\documentclass[%
 reprint,
nofootinbib,
 amsmath,amssymb,
 aps,
]{revtex4-1}

\usepackage{graphicx}
\usepackage{dcolumn}
\usepackage{bm}
\usepackage{hyperref}

\usepackage{braket}
\usepackage{mathtools}
\usepackage{tikz}

\begin{document}

\preprint{APS/123-QED}

\title{
Transmission based tomography for spin qubits 
}

\author{Amritesh Sharma}
 \email{amritesh.iitb@gmail.com}
\author{Ashwin A. Tulapurkar}%
 \email{ashwin@ee.iitb.ac.in}
\affiliation{%
Solid State Devices Group, Department of Electrical Engineering, Indian Institute of Technology, Bombay
}%

\date{\today}

\begin{abstract}
We consider a system of static spin qubits embedded in a one-dimensional spin coherent channel and develop a scheme to readout the state of one and two qubits separately. We use unpolarized flying qubits for this purpose that scatter off from the static qubits due to the Heisenberg exchange interaction. Analysing the transmission coefficient as a function of density matrix elements along with additional unitary gates we reconstruct the state of static qubits.


\end{abstract}

\pacs{Valid PACS appear here}
\maketitle


\section{\label{sec:intro} Introduction}

Measurement of the state of qubits is fundamentally crucial for quantum computing \citep{divincenzo2000physical}. The state (wave function or density operator) of the concerned qubit system is reconstructed from a set of measured observables and is known in literature as state tomography. Since the measurement involves interaction of the well-protected/isolated quantum system with the observer, the state of the system is inevitably perturbed or in some cases collapsed \citep{nielsen2010quantum}. The tomography thus involves simultaneous ensemble measurement of identically prepared states or repeated measurement of a single state prepared identically for each iteration.

There are variety of architecture-oriented tomography techniques. In this paper we are interested in measuring qubits in spintronic systems, especially those where qubits are housed in localized spins (static qubits) embedded in spin-coherent medium. Owing to practical advantages not limited to long decoherence times, potential for all-electrical control and smaller physical footprint (scalability), they offer a promising platform for fault-tolerant quantum information processing\citep{bandyopadhyay2015introduction, he2019two, Ferraro_2020, Watson_2018, Maurand_2016}. Manipulation of qubit states in such systems often relies on local magnetic control \citep{Pla_2012} and or controlled exchange interaction between nearest neighbours \citep{He_2019}. Other approaches found in literature utilize spin-transfer-torque like effects for manipulation of single qubit or entangling multiple qubit states using flying qubits \citep{datta15, Maruri10, Wstate20, Dicke20, YuasaProbeQubit}. From a practical viewpoint local magnetic control is not quite scalable whereas the latter approaches offer indirect access to remote localized qubits via flying qubits. These flying qubits can be provided by a spin-polarized source through mechanisms such as spin-pumping \citep{bhuktare2019direct}, spin-dependent thermoelectric effects \citep{bose2016observation}, spin Hall effect \citep{bose2017sensitive, bose2018observation}, spin Nernst effect \citep{bose2019recent, bose2018direct} an so on. The control of spin qubits via interaction with flying qubits can be compared to its classical counterpart of manipulating nano-magnets \citep{slonczewski1996current, berger1996emission, Shuklaeabc2618}.


An important aspect of the latter mechanism above is the direct exchange interaction between the spin degrees of freedom of the flying and static qubits. The information in the flying qubit is often not utilized for manipulating the information in static qubits \citep{datta15, Wstate20, Dicke20}. The parameters are set such that a single pass (inclusive of multiple scattering inside the channel) of the flying qubits perturbs the combined state of static qubits by minuscule amount in either reflection or transmission channels. The transmission channel was completely blocked to implement unitary operations in those references \citep{datta15, Wstate20, Dicke20}. In this work, however, we shall discuss how we can utilize the information in the flying qubits in the transmission channel to infer the state of static qubits. We shall specifically analyse the transmission of flying qubits through a system of one qubit and two qubits embedded in a spin-coherent channel. We shall also assume the contacts at the ends provide flying qubits and perfectly  accept the flying qubits without any back scattering into the channel.

The idea of using spin-polarized flying qubits for readout of single qubit has been discussed in the literature \citep{YuasaProbeQubit}. As a major distinction from the previous works, we show that the state of single and two qubit system can be measured by using un-polarized flying electrons. We also discuss how the entanglement between static qubits significantly modifies the transmission coefficient. 


\section{\label{sec:model} Model}
In this section we discuss the 1d problem of scattering of spin-polarized electrons by one or two static impurity spins. The static spins are assumed to be non-interacting with each other, while the incident electron and the static spins are assumed to interact via the Heisenberg exchange interaction. Thus if the impurity spin is located at $x=0$, the scattering potential is taken as: $\delta(x) J\bar{\sigma}_f.\bar{\sigma}_s$, where $\bar{\sigma}_f$ and $\bar{\sigma}_s$ denote the Pauli spin-matrix  of incident electron ($f$ stands for flying) and the static impurity respectively, $J$ denotes the Heisenberg exchange interaction strength. To begin with, we consider only one static spin and assume the spin of the static impurity to be frozen along $\hat{n}$ direction. The transmission coefficient ($2 \times 2$ matrix in spin-space) is then given by, 
\begin{equation}
t=[{\mathcal{I}}_2+i\Omega \bm{\hat{n} \cdot \sigma_f}]^{-1}
\end{equation}

where the dimensionless parameter $\Omega$ is given by, $\Omega=mJ/\hbar^2 k$ where m and k denote mass and wave vector of the flying electron. $\mathcal{I}_2$ denotes a $2 \times 2$ identity matrix. The reflection coefficient is related to transmission coefficient as $r=t-\mathcal{I}_2$. The transmission coefficient can be simplified as
 $t=\frac{1}{1+\Omega^2} (\mathcal{I}_2-i\Omega \bm{\hat{n} \cdot \sigma_f})$. One can see that the factor,$t^\dag t$ is given by,  $t^\dag t=\frac{1}{1+\Omega^2} \mathcal{I}_2$.  Thus the transmission probability is independent of the spin direction of the incident electron and is given by $\frac{1}{1+\Omega^2}$. However, the incident spin direction is rotated around n-axis by $\tan^{-1}\frac{2 \Omega}{1-\Omega^2}$ after transmission. Similarly the reflection probability is spin-independent, and the reflected spin is rotated around n-axis by the same angle.

We now extend this calculation to two static spin impurities separated by distance d. We assume that the flying electron interacts via Heisenberg exchange interaction with both the impurities. We can associate a scattering matrix $s=[r \; t'; t \; r']$ with each impurity, where $r$ and $t$ are defined above and $';'$ separates the two rows of the matrix. Further $r'=r$ and $t'=t$ here. The combined s-matrix can be calculated from Ref \citep{Datta_Mesoscopic}. The combined transmission coefficient $t_{comb}$, under these conditions is given by
\begin{equation} \label{eq:tcomb}
t_{comb} = \exp(ikd) t_2 [\mathcal{I}_2-\exp(2ikd) r_1 r_2]^{-1}t_1
\end{equation}
where the subscripts index the corresponding qubit. Taking the first impurity spin along z-direction, and second impurity spin oriented along $(\theta,\phi)$, the factor $t_{comb}^\dag t_{comb}$ turns out to be $\frac{1}{1+2\Omega^2(1+cos\theta)} \mathcal{I}_2$, which is independent of the incident electron spin. Here, we have assumed $kd<<1$ for simplicity. The transmission probability in this case is given by, $P_T=\frac{1}{1+2\Omega^2(1+cos\theta)} $. Thus the transmission probability is again independent of the spin-direction of the incident electron. However the transmission probability depends on the angle between the two static spins. We can consider these two impurities attached to two unpolarized leads, and using the fact that conductance is proportional to the transmission probability, this system shows "magneto-resistance". 
 The rotation of the spin direction of the incident electron can be found from the transmission and reflection coefficients.  
 
 Let's now solve the above two problems assuming the impurity spins to be operators i.e. treating the impurity spins as qubits. In the case of single static impurity, the transmission coefficient ($4 \times 4$ matrix) is 
$t=[{\mathcal{I}_4}+i\Omega \bm{\sigma_f \cdot \sigma_s}]^{-1}$. The reflection coefficient is related to $t$ by $r=t-\mathcal{I}_4$.  t is given by,
\begin{equation} \label{eq:eq1}
t = \frac{1}{1+i \Omega}
\begin{bmatrix}
    1      & 0      & 0  &0       \\
    0      & \frac{\Omega+i}{3\Omega+i} & \frac{2\Omega}{3\Omega+i} &0       \\
    0 &\frac{2\Omega}{3\Omega+i} & \frac{\Omega+i}{3\Omega+i} &0\\
    0&0&0&1
\end{bmatrix} 
\end{equation}
The transmission and reflection probabilities are given by $P_T=\text{trace}(t^\dag t \rho)$ and $P_R=\text{trace}(r^\dag r \rho)$, where $\rho$ is the density matrix of the combined system of incident electron (flying qubit) and static spin impurity (static spin qubit). Assuming that initially the flying qubit is polarized along z direction and static qubit  along $(\theta,\phi)$ direction, the transmission probability turns out to be,
$P_T=\frac{(7\Omega^2+1)+2\Omega^2 cos(\theta)}{(\Omega^2+1)(9\Omega^2+1)}$.
Note that the transmission probability is independent of the sign of $\Omega$. The transmission probability is $\theta$ dependent, and thus we get "magneto-resistance" if the impurity spin is treated as operator. After the scattering the system is entangled, and the rotation of the spin directions can be found from the transmission and reflection coefficients. The spin of the flying electron develops a spin-polarization along the spin direction of static qubit and there are also changes in the spin-polarization in the transverse direction. The same also holds for the static qubit's spin i.e. it develops a spin-polarization along the spin direction of flying qubit along with changes in the transverse polarization.   

\begin{figure}
\includegraphics[width=0.48\textwidth, trim={0 0 0 0},clip]{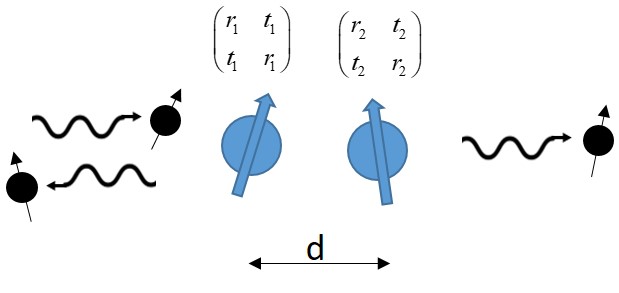}
\caption[]{\textbf{scattering of flying qubit by two static qubits}.  Individual qubits act as spin-dependent scatterers with reflection and transmission denoted by $[r]$ and $[t]$ matrices. }
\label{fig:figl}
\end{figure}

We now consider the case of two static qubits. We assume that the static qubits are non-interacting with each other. The flying qubit interacts with each static qubit via Heisenberg exchange interaction. We can associate $8 \times 8$ scattering matrix with each static qubit, $s=[r \; t'; t \; r']$, and combine the two s-matrices to get combined s-matrix of the system, from which we can find out the transmission and reflection coefficients (see Fig.~\ref{fig:figl}). Knowing the initial  density matrix of the combined system of flying and two static qubits, we can find out the transmission probability as $P_T=\text{trace}(t^\dag_{comb} t_{comb} \, \rho_i)$, where $t_{comb}$ is the combined transmission matrix and $\rho_i$ is the initial density matrix. Below we give the expression for transmission probability for the case where the incident flying qubit is un-polarized and the static qubits are described by a density matrix $(\rho)$, which was obtained after some algebraic manipulations.
\begin{equation}
P_T=\\
\frac{(1+12\Omega^2)+4\Omega^2(1+8\Omega^2)(\rho_{22}+\rho_{33}-2Re(\rho_{23}))}{(1+16\Omega^2)(1+4\Omega^2)}
\end{equation}
Note that we have assumed that $kd<<1$ in writing above equation, where $d$ is the distance between the static qubits. The above equation is valid even for a mixed state density matrix of static qubits. As the density matrix is non-negative definite matrix, $\rho_{22}+\rho_{33}-2Re(\rho_{23})\geq 0$, which makes sure that transmission probability is non-negative for any value of $\Omega$. The factor $\rho_{22}+\rho_{33}-2Re(\rho_{23})$ can not exceed 2 as the transmission probability can not be more than 1. In fact, when this factor is 2, the transmission probability is 1, for any value of $\Omega.$ As discussed in the appendix, the above equation can be written in a physically transparent way by noting that, $\frac{1-\langle \bm{\sigma_1 \cdot \sigma_2} \rangle}{2}= \rho_{22}+\rho_{33}-2Re(\rho_{23}$), where $\bm{\sigma_1}$ and $\bm{\sigma_2}$ are Pauli spin operators of first and second static qubit respectively, and $\langle \rangle$ denote average value. Further, we can write, $\frac{1-\langle \bm{\sigma_1 \cdot \sigma_2} \rangle}{2}=2-\frac{\langle \bm{S}^2 \rangle}{\hbar^2}$, where $\bm{S}$ is the total spin operator of the two static qubits.
 Note that the maximum and minimum value of  $\langle \bm{S}^2/\hbar^2 \rangle$ is 2 and 0 corresponding to the triplet and singlet states respectively. Thus the transmission is maximum (equal to 1, for any value of $\Omega$) for the singlet state and minimum for the triplet state. 
Consider a simple case where the first static qubit is polarized along z-axis and the second qubit along $(\theta,\phi)$ direction. As the qubits are un-entangled, $\langle \bm{\sigma_1 \cdot \sigma_2} \rangle=\langle \bm{\sigma_1}\rangle \cdot \langle \bm{\sigma_2} \rangle=\cos(\theta)$.  We thus see that the transmission depends on $\cos(\theta)$, but the functional form is quite different compared to the previous case where the two impurity spins were considered to be frozen. If the static qubit spins are un-entangled, $\langle \bm{\sigma_1 \cdot \sigma_2} \rangle$ ranges from 1 to -1. In general case, the value of $\langle \bm{\sigma_1 \cdot \sigma_2} \rangle$ ranges from 1 (triplet state) to -3 (singlet state). Thus the transmission probability can be significantly increased due to the entanglement. 

\section{\label{sec:Tomography} Tomography Schemes}
\subsection{Tomography of Single Qubit}
We now examine how tomography of single and two qubit system can be performed from the measurement of transmission probability using un-polarized electrons.
Consider a single static impurity qubit with density matrix is to be determined. We assume that multiple copies of the impurity qubit are available. To determine the density matrix, we place an ancilla qubit near the  impurity qubit and carry out the transmission measurements using un-polarized flying qubits. If the ancilla qubit is polarized along z-direction, we can measure the average value, $\langle \bm{\sigma_z} \rangle$ of the impurity qubit. By changing the polarization of ancilla qubit to x and y directions, we can measure  $\langle \bm{\sigma_x} \rangle$ and $\langle \bm{\sigma_y} \rangle$ values. From these three average values, the density matrix can be determined as, discussed in the appendix. Instead of rotating the ancilla qubit, one can also rotate the impurity qubit by using single qubit gates.

\subsection{Tomography of two qubit system}
We now discuss how tomography of two-qubit system can be carried out by measuring the transmission probability of incident un-polarized flying qubit. As discussed in the Appendix.~\ref{sec:appendix}, to know the density matrix completely we need to find out 15 $a$ coefficients which correspond to to average values of certain operators. We thus need 15 equations relating the $a$ coefficients. We have seen in the previous section that measurement of transmission probability depends on $\langle \bm{\sigma_1 \cdot \sigma_2} \rangle$ i.e. it gives us the value of $(a_{1,1}+a_{2,2}+a_{3,3})$. We now need two more equations relating $a_{1,1}$, $a_{2,2}$ and $a_{3,3}$ to determine them. We now apply certain gates (unitary operators, U) to the qubit system and measure the transmission probability in the new state ($\rho_{new}=U \rho U^\dag$). The gates are chosen to give us the required two equations. Consider applying single qubit X gate to the second qubit. This corresponds to $U=\mathcal{I}_2 \otimes \sigma_x$. One can easily check that $U M_{1,1} U^\dag =M_{1,1}$, $U M_{2,2} U^\dag =-M_{2,2}$, and $U M_{3,3} U^\dag =-M_{3,3}$, where the various $M$ matrices are defined in the appendix. Thus the values of $a_{2,2}$ and $a_{3,3}$ in the new state change sign. Thus the transmission probability in the new state gives us value of $(a_{1,1}-a_{2,2}-a_{3,3})$. Similarly measurement of transmission probability of the state obtained from the original state after application of Y gate to the second qubit, gives us $(-a_{1,1}+a_{2,2}-a_{3,3})$. From these three equations we can obtain $a_{1,1}$, $a_{2,2}$ and $a_{3,3}$. Instead of applying single qubit gates to the second qubit, we can as well apply them to the first qubit and will get the same information.

Let's now see other coefficients can be measured. Consider  single qubit rotation around y-axis by $\pi/2$. This essentially changes z into x and x into -z. Thus taking $U=\mathcal{I}_2 \otimes R_y(\pi/2)$, we get $U M_{1,3} U^\dag =M_{1,1}$, $U M_{2,2} U^\dag =M_{2,2}$ and $U M_{3,1} U^\dag =-M_{3,3}$ i.e. $a_{1,1,new}=a_{1,3}$, $a_{2,2,new}=a_{2,2}$ and $a_{3,3,new}=-a_{3,1}$. Thus measurement of transmission probability gives value of $a_{1,3}-a_{3,1}$, as value of $a_{2,2}$ is known. Now consider single qubit gate $X R_y(\pi/2)$. Note that X gate changes y into -y and z into -z. Thus the combined operator $X R_y(\pi/2)$ changes x into z and and z into x (and y into -y). Thus by applying  $U=\mathcal{I}_2 \otimes X R_y(\pi/2)$ to the original state and measuring the transmission probability we get the value of $a_{1,3}+a_{3,1}$. From the two equations for $a_{1,3}$ and $a_{3,1}$ we can determine them. (Note that $X R_y(\pi/2)$ is same as Hadamard gate). Using this method coefficients $a_{1,2}$ and $a_{2,1}$ can be determined by application of $R_z(\pi/2)$ and $Y R_z(\pi/2)$ single qubit gates. Similarly, coefficients $a_{2,3}$ and $a_{3,2}$ can be determined by application of $R_x(\pi/2)$ and $Z R_x(\pi/2)$ single qubit gates.   

We still need to determine six more coefficients: $a_{0,i}$ and $a_{i,0}$ where i = 1,2 or 3. These can not be determined by applying single qubit gates as index 0 can not be converted into other non-zero indices by these gates. we need to apply two qubit gates to get the remaining six $a$ coefficients. We can choose square root SWAP gate as a two qubit gate. The average $\langle \bm{\sigma_1 \cdot \sigma_2} \rangle$ is invariant under this operation. However this gate does change $M_{0,i}$ and $M_{i,0}$ matrices e.g. $M_{3,2,new}=M_{0,1}+M_{1,0}$.  If we apply single qubit $R_x(\pi/2)$ gate on the second qubit after applying sqrt(SWAP) gate, the resulting transmission probability depends on $a_{0,1}$ and $a_{1,0}$. Applying $R_z$ on qubit 2, followed by sqrt(SWAP), followed by $R_x(\pi/2)$ on qubit 2 gives one more equation for $a_{0,1}$ and $a_{1,0}$. Thus from these two operations, $a_{0,1}$ and $a_{1,0}$ can be determined. In a similar way we can determine $a_{0,2}, a_{2,0}$ and $a_{0,3},a_{3,0}$. Note that the in principle there are many different choices of single and two qubit gates. We have selected here some of the "standard" gates.

\begin{figure}
\includegraphics[width=0.48\textwidth, trim={0 0 0 0},clip]{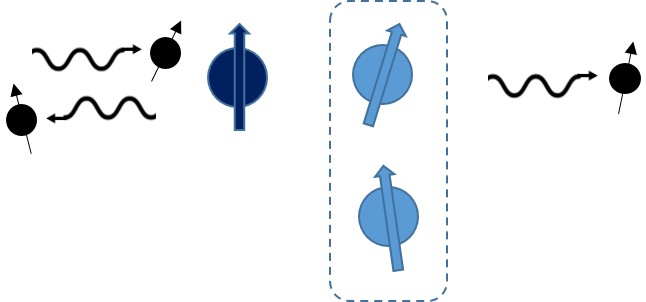}
\caption[]{\textbf{Measurement of the average spin values of the first qubit}.  The two static impurity spins are denoted by blue colour. An ancilla qubit is placed near the first static qubit. Flying qubit is passed through the ancilla qubit and the first static qubit.  }
\label{fig:fig2}
\end{figure}

It is possible to avoid usage of two qubit gates in the above scheme. Note that six coefficients: $a_{i,0}$ and $a_{0,i}$ with $i = 1,2$ or 3, essentially are the average values of the $\sigma_x$, $\sigma_y$ and $\sigma_z$ of the first and second qubit respectively. The average values for the first qubit  can be measured as shown in  Fig.~\ref{fig:fig2}. If the flying qubit interacts only with the first qubit, the transmission probability depends on the average value of $\sigma$ of the first qubit along the ancilla qubit direction. Thus by varying the polarization of the ancilla qubit along x,y and z directions we can measure average values of $\sigma_x$, $\sigma_y$ and $\sigma_z$ of the first qubit. This gives us the coefficients $a_{1,0}$, $a_{2,0}$ and $a_{3,0}$. Similarly we can carry out the measurements on the second qubit to get $a_{0,1}$, $a_{0,2}$ and $a_{0,3}$.
Note that in the type of measurements as shown in Fig.~\ref{fig:fig2}, what matters is the density matrix traced over the second qubit. It shown in the appendix that this operation results in density matrix which depends on the average values of Pauli operators of the first qubit. As a related comment, it should be noted that any single qubit unitary operation performed on the second qubit does not change the transmission probability through the first qubit, even if the qubits are entangled. It is thus not possible to modulate the transmission via entanglement. If we carry out a projective measurement on the second qubit, the state of the first qubit will collapse, but the transmission on average remains the same.

{\bf Tomography using polarized detectors/injectors:} Up to now we have considered un-polarized incident flying qubits and looked at the measurement of transmission probability. If we can also measure the spin-polarization of the transmitted flying qubit, we can get additional information. If the flying qubit is transmitted, the density matrix of the system is given by : $t_{comb} \rho_i t_{comb}^\dag/P_T$. The average value of the flying qubit's Pauli spin operators obtained from the density matrix is given by: 
$\frac{6 \Omega^2}{(1+16\Omega^2) (1+4\Omega^2)} \frac{\langle \bm{\sigma_1}+\bm{\sigma_2} \rangle}{P_T}$. Thus the incident un-polarized flying qubit gets polarized after transmission and the average spin-polarization is along the net spin direction of the static qubits. Thus from these measurements, we can measure $\langle \bm{\sigma_1} \cdot \bm{\sigma_2} \rangle$ and $\langle \bm{\sigma_1}+\bm{\sigma_2} \rangle$. In terms of $a$ coefficients, the measurement of x,y and z components of  $\langle \bm{\sigma_1}+\bm{\sigma_2} \rangle$ corresponds to measuring  $a_{1,0}+a_{0,1}$, $a_{2,0}+a_{0,2}$ and $a_{3,0}+a_{0,3}$ respectively.  If we apply X gate to the second qubit, the average value of $\langle \sigma_{2,y} \rangle$ and $\langle \sigma_{2,z} \rangle$ changes sign, and we can measure values of $a_{2,0}-a_{0,2}$ and $a_{3,0}+a_{0,3}$. If we apply Y gate to the second qubit, the average value of $\langle \sigma_{2,x} \rangle$ and $\langle \sigma_{2,z} \rangle$ changes sign, and we can measure values of $a_{1,0}-a_{0,1}$ and $a_{3,0}+a_{0,3}$. Thus we can measure  $a_{0,123}$ and $a_{123,0}$ without need of two qubit gates or the arrangement as shown in Fig.~\ref{fig:fig2}.

Instead of using un-polarized incident electron and detecting the polarization of the transmitted electron, we can use polarized electron as input and then measure the transmission probability. If the incident electron is polarized along z direction, the transmission probability is given by:  $P_T -\frac{2 \Omega^2}{(1+16\Omega^2)(1+4\Omega^2)} (\sigma_{1,z}+\sigma_{2,z})$, where $P_T$ is the transmission probability of un-polarized electron. Thus measuring the transmission probability with electrons polarized along and opposite to x,y and z directions gives us the values of   $\langle \bm{\sigma_1} \cdot \bm{\sigma_2} \rangle$ and $\langle \bm{\sigma_1}+\bm{\sigma_2} \rangle$.

{\bf Tomography of pure state:} If the two static qubits are in a pure state, we can write the wavefunction as:
$\ket{\psi}=a_1 e^{i \theta_1} \ket{00}+ a_2 e^{i \theta_2} [\ket{01}+\ket{10}/\sqrt{2}]+ a_3 [\ket{01}-\ket{10}/\sqrt{2}] +a_4 e^{i \theta_4} \ket{11}$. We have chosen $\theta_3$ to be 0. There are six unknown parameters due to the normalization condition. We can see that $\langle \bm{\sigma_1} \cdot \bm{\sigma_2} \rangle=1-4a_3^2$. Thus measurement of the transmission probability of un-polarized electron gives us $a_3$ parameter. As done previously, we can apply various gates to the static qubits and measure the transmission probability. If we measure transmission probability after applying X, Y and Z gates to the second qubit, we can get amplitudes of all the four wavefunction components i.e. $a_1$, $a_2$, $a_3$ and $a_4$.   We can apply $R_x(\pi/2)$, $R_y(\pi/2)$ and $R_z(\pi/2)$ gates to the second qubit and measure transmission probability. This gives us partial information about the phases. e.g. we get value of $sin(\theta_2)$ leaving an uncertainty of $\pi-\theta_2$. If we measure transmission probability with polarized qubits, such an uncertainty can be removed.

We have assumed Heisenberg exchange interaction between the flying and each static qubit. If the interaction Hamiltonian is invariant under any unitary operation U, the transmission coefficients $t_1, t_2$ and hence the combined s- matrix are also invariant. This implies that if the initial density matrix is transformed under U, the transmission probability remains the same. The interaction Hamiltonian here is invariant under rotation of all spins. The previous results viz. transmission probability of un-polarized electrons depends on $\langle \bm{\sigma_1} \cdot \bm{\sigma_2} \rangle$ and transmission probability of spins polarized along $\hat{n}$ depend on $\langle \bm{\sigma_1} \cdot \bm{\sigma_2} \rangle$ and $\langle (\bm{\sigma_1} + \bm{\sigma_2}) \cdot \hat{n} \rangle$ are consistent with these symmetry arguments. (It should be also noted that the transmission is invariant under time reversal.) We have assumed the parameter $kd$ to be small, which could be $2n\pi$ in principle, while combining the s-matrices. For non-zero distances d, the transmission depends on the parameter $\exp(ikd)$ (see Eq.~\ref{eq:tcomb}). However, this does not change the qualitative nature of transmission probability i.e. it still depends on $\langle \bm{\sigma_1} \cdot \bm{\sigma_2} \rangle$ and $\langle (\bm{\sigma_1} + \bm{\sigma_2}) \cdot \hat{n} \rangle$. The coefficients of various terms depend on $\exp(ikd)$ factor in a complicated way. As an example, the previous result that transmission is 1 for singlet state no longer holds. This results is valid only for $kd=2n\pi$.  However the previous algorithms for tomography would still work as the transmission still contains information about $\langle \bm{\sigma_1} \cdot \bm{\sigma_2} \rangle$ and $\langle (\bm{\sigma_1} + \bm{\sigma_2}) \cdot \hat{n} \rangle$. This argument can be extended to multiple qubit system e.g. in the case of three qubit system, the transmission probability would depend on $\langle \bm{\sigma_1} \cdot \bm{\sigma_2}+\bm{\sigma_2} \cdot \bm{\sigma_3}+\bm{\sigma_3} \cdot \bm{\sigma_1} \rangle$ and and $\langle (\bm{\sigma_1} + \bm{\sigma_2}+ \bm{\sigma_3}) \cdot \hat{n} \rangle$. In this way, the present scheme can be extended to tomography of multiple qubit system.

We now consider optimal values of the parameters ($kd$  and $\Omega$) for the case of tomography by un-polarized flying qubits. If $kd=2n\pi$, the transmission coefficient ($P_T$) is one for singlet state and $(1+12\Omega^2)/ (1+16 \Omega^2)(1+ 4\Omega^2)$ for triplet state. Thus a larger value of $\Omega$ would give a larger variation in the transmission coefficient. In the limit of large $\Omega$, the $P_T$ for triplet state would be zero. As the $P_T$ values are at maximum and minimum possible values, the first derivative of $P_T$ w.r.t. the parameters $kd$  and $\Omega$ is zero implying that such a choice would be robust against variations in the parameters.

\begin{figure}
\includegraphics[width=0.48\textwidth, trim={0 0 0 0},clip]{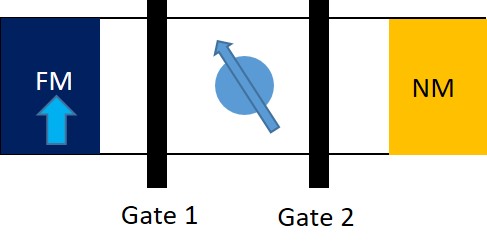}
\caption[]{\textbf{Heat Engine}.  A static qubit (light blue colour) is placed between a ferromagnetic (FM) and non-magnetic (NM) reservoir. By opening gate 1 and closing gate 2, the qubit gets polarized along the magnetization direction of the FM. By opening gate 2 and closing gate 1, the qubit gets de-polarized.}
\label{fig:fig3}
\end{figure}

We finally consider another application of the system of flying and static qubits. Consider a static qubit placed between a polarized (FM) and an un-polarized (NM) reservoir as shown in Fig.~\ref{fig:fig3}. There are two controllable gates between the reservoirs and static qubit. If we keep the gate 2 closed and gate 1 open, the static qubit interacts with the polarized flying qubits incident from the polarized reservoir. The flying qubits finally return to the reservoir as the gate 2 acts like a perfect reflector. The flying and static qubit get entangled due to the Heisenberg exchange interaction, and the state of the static qubit is changed when the flying qubit returns to the reservoir which corresponds to taking partial trace of the combined density matrix over the flying qubit. This effect has been analyzed in Ref. \citep{datta15}, and subsequently in Ref. \citep{Wstate20, Dicke20}. After sequential interaction with many flying qubits, the static qubit gets polarized along the polarization direction of the reservoir. We now close the gate 1 and open gate 2. The static qubit now interacts with un-polarized flying qubits emerging from the second reservoir. After interaction with many flying qubits, the static qubit gets un-polarized. After interaction with the first reservoir, the static qubit's state is a pure state with zero entropy and after interaction with the second reservoir, the static qubit is in a completely mixed state with entropy $k_B ln(2)$. Thus in one cycle a maximum entropy of $k_B ln(2)$ can be transferred from the un-polarized reservoir to the polarized reservoir. Thus the system shown in Fig.~\ref{fig:fig3} can work as a heat engine. We can replace the single qubit by many non-interacting static qubits. Our numerical simulations indicate that it is possible to "magnetize" and "de-magnetize" the qubits by connecting them to the ferromagnetic and non-magnetic reservoirs. The process however needs some single qubit operations on the qubits (which do not change entropy) to completely "magnetize" and "de-magnetize" the qubits. 

\section{\label{sec:conclusion} Conclusion}
In this paper, we have analysed the transmission of flying qubits from a system of static impurity spins. 


The idea of 'quantum magneto-resistance' is introduced whereby the transmission probability and hence the conductance depends on the entangled quantum state of the  the static qubits.  We explicitly obtained the expressions of transmission probability as a function of density matrix components for one or two qubits. The tomography scheme we develop hinges on the fact that transmission probability through a two qubit system  depends on the expectation value of the scalar product of spin operators of the two qubits. Measurement of the transmission coefficient after application of appropriate unitary gates is sufficient for inferring the density matrix. For tomography of a single static qubit, we use another ancilla static qubit in a known state.
Finally, we discuss another similar scenario containing a single qubit connected to two different types of reservoirs (one polarized and other un-polarized) each with controlled gates but now utilizing only the reflection channels. Alternating connections to these reservoirs enables transfer of entropy which are the tell-tale signs of a minuscule heat engine.



\appendix 

\section{\label{sec:appendix}}

The density matrix for a single qubit can be resolved as:
\begin{equation} \label{eq:density_mat_1qubit}
\rho=\frac{1}{2} [\mathcal{I}_2+ \langle \sigma_x \rangle \sigma_x+ \langle \sigma_y \rangle \sigma_y +\langle \sigma_z \rangle \sigma_z]
\end{equation}
where average of an operator $O$ is given by, $\langle O \rangle=trace(\rho O)$. Thus density matrix is uniquely determined if we know $\langle \sigma_x \rangle$, $\langle \sigma_y \rangle$ and $\langle \sigma_z \rangle$.

For two qubits, the density matrix ($4 \times 4$) can be resolved into 16 matrices obtained from the set $(\mathcal{I}_2,\sigma_x,\sigma_y, \sigma_z) \otimes (\mathcal{I}_2,\sigma_x,\sigma_y, \sigma_z)$. We can write,

\begin{equation} \label{eq:density_mat_2qubits}
\begin{aligned}
\rho=\frac{1}{4}[a_{0,0}(\mathcal{I}_2 \otimes \mathcal{I}_2)+a_{0,1} (\mathcal{I}_2 \otimes \sigma_x)+ a_{0,2} (\mathcal{I}_2 \otimes \sigma_y)\\+a_{0,3} (\mathcal{I}_2 \otimes \sigma_z)+a_{1,0} (\sigma_x \otimes \mathcal{I}_2) +a_{1,1} (\sigma_x \otimes \sigma_x) + \\ a_{1,2} (\sigma_x \otimes \sigma_y)+...+a_{3,3} (\sigma_z X \sigma_z)]
\end{aligned}
\end{equation} 

All the $a$ coefficients are real and given by $a_{i,j}=\langle  \sigma_i \otimes \sigma_j \rangle$, where $\sigma_0=\mathcal{I}_2, \sigma_1=\sigma_x$ etc. Note that $a_{0,0}=1$ as trace of density matrix is 1. We will denote the various matrices in the above equation by symbol $M$ i.e. $M_{i,j}=\sigma_i \otimes \sigma_j$. To find out the density matrix of two qubit system, we need to know these 15 coefficients. 
Taking the above form of density matrix, we find that $\rho_{22}+\rho_{33}-2Re(\rho_{23})=\frac{1}{2}[1-(a_{1,1}+a_{2,2}+a_{3,3})]$. Further using, $a_{1,1}+a_{2,2}+a_{3,3}=\langle \bm{\sigma_1 \cdot \sigma_2} \rangle$, we see that the transmission is determined by $\langle \bm{\sigma_1 \cdot \sigma_2} \rangle$. Thus we can write,
\begin{equation}
P_T=\\
\frac{(1+12\Omega^2)+2\Omega^2(1+8\Omega^2)(1-\langle \bm{\sigma_1 \cdot \sigma_2} \rangle)}{(1+16\Omega^2)(1+4\Omega^2)}
\end{equation}

If we take the partial trace of the density matrix over the second qubit, we get,
\begin{equation} \label{eq:eq1}
\rho_1 = \frac{1}{2}[ \mathcal{I}_2-
\begin{bmatrix}
    a_{3,0}      & a_{1,0}-i a_{2,0}       \\
    a_{1,0}+i a_{2,0}      &-a_{3,0} 
\end{bmatrix} 
]
\end{equation}
The traced out matrix can be compared to equation \ref{eq:density_mat_1qubit}. It shows that the average values of the Pauli operators of the first qubit in the combined density matrix are the same as average values obtained from the traced out density matrix as well. 

If we take the partial trace of the density matrix over the first qubit, we get,
\begin{equation} \label{eq:eq1}
\rho_2 = \frac{1}{2}[ \mathcal{I}_2-
\begin{bmatrix}
    a_{0,3}      & a_{0,1}-i a_{0,2}       \\
    a_{0,1}+i a_{0,2}      &-a_{0,3} 
\end{bmatrix} 
]
\end{equation}

\begin{acknowledgments}
We acknowledge the support of Department of Science and Technology (DST), Government of India through Project No. SR/NM/NS-1112/2016 and Science and Engineering Research Board (SERB) through Project No. EMR/2016/007131.
\end{acknowledgments}


\bibliography{QMR_v7}

\end{document}